\documentclass[12pt]{iopart}

\usepackage{bm}
\usepackage{amssymb}
\usepackage{latexsym}
\usepackage{amsfonts}
\usepackage{epsfig}
\usepackage{psfrag}

\usepackage{color}

\newcommand{\nn}{\nonumber\\}

\newcommand{\f}[1]{\mbox{\boldmath$#1$}}

\newcommand{\bea}{\begin{eqnarray}}
\newcommand{\eea}{\end{eqnarray}}
\newcommand{\ord}{{\cal O}}

\newcommand{\trace}[1]{{\rm Tr}\left\{ #1 \right\}}
\newcommand{\ii}{{\rm i}}

\begin{document}

\title{Duality in spin systems via the $SU(4)$ algebra}

\author{Gernot Schaller}
\ead{gernot.schaller@tu-berlin.de}
\address{Institut f\"ur Theoretische Physik, Hardenbergstra{\ss}e 36,
Technische Universit\"at Berlin, D-10623 Berlin, Germany}
\author{Ralf Sch\"utzhold}
\ead{schuetz@theo.physik.uni-due.de}
\address{Institut f\"ur Theoretische Physik, Lotharstra{\ss}e 1,
Universit\"at Duisburg-Essen, D-47048 Duisburg, Germany}

\begin{abstract}
We provide several examples and an intuitive diagrammatic representation 
demonstrating the use of two-qubit unitary transformations for mapping coupled 
spin Hamiltonians to simpler ones and {\em vice versa}.
The corresponding dualities may be exploited to identify phase transition 
points or to aid the diagonalization of such Hamiltonians.
For example, our method shows that a suitable one-parameter family of 
coupled Hamiltonians whose ground states transform from an initially 
factorizing state to a final cluster state on a lattice of arbitrary dimension 
is dual to a family of trivial decoupled Hamiltonians containing local on-site terms only. 
As a consequence, the minimum enery gap 
(which determines the adiabatic run-time) does not scale with system size, 
which facilitates an efficient and simple adiabatic preparation of 
e.g.\ the two-dimensional cluster state used for measurement-based quantum computation.
\end{abstract}

\pacs{64.70.Tg, 
03.65.Ud 
05.30.Rt, 
}

\maketitle

\section{Introduction}

In order to diagonalize a nontrivial Hamiltonian, it is often mapped via 
a (unitary) similiarity transformation to a decoupled one, which consists 
of a sum of operators locally acting on distinct parts.
The advantages of this procedure are obvious: 
Decoupled Hamiltonians may be diagonalized readily when the size of their 
local parts is small (e.g., only one qubit or spin-1/2).
The spectrum is unaffected by the similiarity transformation and the 
eigenvectors of the original Hamiltonian may be obtained by applying the 
inverse similarity transformation to the simple eigenvectors of the 
decoupled Hamiltonian.

Single-qubit rotations can be described by the symmetry group $SU(2)$, 
which is intimately related to the group of three-dimensional rotations 
$SO(3)$.
Our intuitive understanding of the latter makes local duality
transformations such as 
\bea
U\left[\sum_i g_i \sigma^x_i+ J_i \sigma^z_i\sigma^z_{i+1}\right]U^\dagger
=
\sum_i g_i \sigma^z_i+J_i \sigma^x_i\sigma^x_{i+1}
\eea
quite simple to follow.
In contrast, general two-qubit transformations 
$SU(4) \simeq SU(2) \otimes SU(2)$ lead to less obvious dualities.
However, they are of significant interest, since all unitary transformations 
on an $n$-qubit system may be expressed by products of 
(in the worst case $\ord\left\{2^n\right\}$) two-qubit transformations.
E.g., the active field of quantum information has essentially been so 
attractive since some unitary transformations 
(as the quantum Fourier transform) may be expressed by a small 
(polynomial in $n$) number of two-qubit operations only~\cite{nielsen2000}.

Here, we will follow a different objective: 
Instead of trying to find an optimal quantum circuit for computation, 
we would like to find an optimal unitary that decouples a given Hamiltonian.
For simplicity, we will constrain ourselves to two-qubit rotations only.
More general unitaries can of course be constructed from products of 
two-qubit rotations, it is however also conceivable to construct them 
directly from the generators of $SU(N)$.

\section{$SU(4)$ properties and Notation}

The Hilbert space of two qubits is by construction four-dimensional.
Therefore, all linear operators acting on this Hilbertspace can be formed 
by complex linear combinations of 16 basis matrices, which we can choose 
to be hermitian and trace-orthogonal.
These basis matrices can be easily constructed from the direct product of 
the identity matrix and the Pauli matrices acting on either subspace, 
i.e., with choosing
\bea
\label{Egenerators}
\Sigma^{\alpha\beta} &=& \frac{1}{2} \sigma^\alpha_1 \otimes \sigma^\beta_2
\,,
\eea
where $\alpha,\beta \in \{\circ,x,y,z\}$ and $\sigma^{\circ}\equiv \f{1}$ 
denotes a two by two identity matrix, 
we automatically obtain a hermitian basis which satisfies trace orthogonality
\mbox{$\trace{\Sigma^{\alpha\beta} \Sigma^{\gamma\delta}}=
\delta_{\alpha\gamma}\delta_{\beta\delta}$}.
Similarly, the generators of all $SU(N)$ may be (recursively) 
related to the generators of the factors
of $SU(N_1)$ and $SU(N_2)$, where $N=N_1 N_2$.

For $SU(4)$, it is evident that $\Sigma^{\circ\circ}$ will only give rise to a 
global phase.
Furthermore, it is evident that $\Sigma^{\circ x},\Sigma^{\circ y},\Sigma^{\circ z}$ 
act non-trivially only on the second (right) qubit and 
similarly $\Sigma^{x\circ},\Sigma^{y\circ},\Sigma^{z\circ}$ only on the first 
(left) qubit.
These transformations correspond to local rotations of the first or second 
qubit only, respectively, 
and cannot transform local and non-local terms into each other. 
This is different for rotations from $\Sigma^{xx}$ to $\Sigma^{zz}$, 
which are in the focus of our present paper.

The most general unitary operation on 2 qubits can be parameterized by 
16 real parameters, i.e.,
\bea
U = \exp\left\{\ii \sum_{\alpha,\beta\in\{\circ,x,y,z\}} 
a_{\alpha\beta} \Sigma^{\alpha\beta}\right\}\,,
\eea
where $a_{\alpha\beta} \in \mathbb{R}$.
If we are not interested in a global phase, we can set $a_{\circ\circ}=0$.
Prominent cases are for example the controlled NOT gate
\bea\label{Egatecx}
S^{\rm CX} &=& 
\exp\left\{\ii 
\frac{\pi}{2} 
\left[
\Sigma^{\circ \circ}-\Sigma^{\circ x}-\Sigma^{z \circ}+\Sigma^{z x}\right]
\right\}\nn
&=& \Sigma^{\circ \circ}+\Sigma^{\circ x}+\Sigma^{z \circ}-\Sigma^{z x}\,,
\eea
the controlled Z gate
\bea\label{Egatecz}
S^{\rm CZ} &=&
\exp\left\{\ii 
\frac{\pi}{2} 
\left[
\Sigma^{\circ \circ}-\Sigma^{\circ z}-\Sigma^{z \circ}+\Sigma^{z z}
\right]\right\}\nn
&=& \Sigma^{\circ \circ}+\Sigma^{\circ z}+\Sigma^{z \circ}-\Sigma^{z z}\,,
\eea
and the SWAP gate
\bea\label{Egateswap}
S^{\rm SWAP} &=& 
\exp\left\{\ii 
\frac{\pi}{2} 
\left[
-\Sigma^{\circ \circ}+\Sigma^{x x}+\Sigma^{y y}+\Sigma^{z z}
\right]\right\}\nn
&=& \Sigma^{\circ \circ}+\Sigma^{x x}+\Sigma^{y y}+\Sigma^{z z}\,.
\eea

{For an ordinary $SU(2)$ rotation, all axes except the rotation axis will be modified.
In contrast, when}
considering rotations around a single axis $\Sigma^{\alpha\beta}$ only 
by an angle $\eta \in \mathbb{R}$, i.e., 
$U^{\alpha\beta} = \exp\left(\ii \eta \Sigma^{\alpha\beta}\right)$, 
one observes that these rotate only four pairs of axes into each other while 
keeping the remaining ones invariant, see table~\ref{Trotations}.
\begin{table}[ht]
\begin{tabular}{c|c|c|c|c}
axis & pair 1 & pair 2 & pair 3 & pair 4\\
\hline
$\Sigma^{\circ x}$ & 
$\Sigma^{\circ y} \stackrel{(-)}{\leftrightarrow} \Sigma^{\circ z}$ & $\Sigma^{x y} \stackrel{(-)}{\leftrightarrow} \Sigma^{x z}$ &
$\Sigma^{y y} \stackrel{(-)}{\leftrightarrow} \Sigma^{y z}$ & $\Sigma^{z y} \stackrel{(-)}{\leftrightarrow} \Sigma^{z z}$\nn
$\Sigma^{\circ y}$ &
$\Sigma^{\circ x} \stackrel{(+)}{\leftrightarrow} \Sigma^{\circ z}$ & $\Sigma^{x x} \stackrel{(+)}{\leftrightarrow} \Sigma^{x z}$ &
$\Sigma^{y x} \stackrel{(+)}{\leftrightarrow} \Sigma^{y z}$ & $\Sigma^{z x} \stackrel{(+)}{\leftrightarrow} \Sigma^{z z}$\nn
$\Sigma^{\circ z}$ &
$\Sigma^{\circ x} \stackrel{(-)}{\leftrightarrow} \Sigma^{\circ y}$ & $\Sigma^{x x} \stackrel{(-)}{\leftrightarrow} \Sigma^{x y}$ &
$\Sigma^{y x} \stackrel{(-)}{\leftrightarrow} \Sigma^{y y}$ & $\Sigma^{z x} \stackrel{(-)}{\leftrightarrow} \Sigma^{z y}$\nn
\hline
$\Sigma^{x \circ}$ &
$\Sigma^{y \circ} \stackrel{(-)}{\leftrightarrow} \Sigma^{z \circ}$ & $\Sigma^{y x} \stackrel{(-)}{\leftrightarrow} \Sigma^{z x}$ &
$\Sigma^{y y} \stackrel{(-)}{\leftrightarrow} \Sigma^{z y}$ & $\Sigma^{y z} \stackrel{(-)}{\leftrightarrow} \Sigma^{z z}$\nn
$\Sigma^{y \circ}$ &
$\Sigma^{x \circ} \stackrel{(+)}{\leftrightarrow} \Sigma^{z \circ}$ & $\Sigma^{x x} \stackrel{(+)}{\leftrightarrow} \Sigma^{z x}$ &
$\Sigma^{x y} \stackrel{(+)}{\leftrightarrow} \Sigma^{z y}$ & $\Sigma^{x z} \stackrel{(+)}{\leftrightarrow} \Sigma^{z z}$\nn
$\Sigma^{z \circ}$ &
$\Sigma^{x \circ} \stackrel{(-)}{\leftrightarrow} \Sigma^{y \circ}$ & $\Sigma^{x x} \stackrel{(-)}{\leftrightarrow} \Sigma^{y x}$ &
$\Sigma^{x y} \stackrel{(-)}{\leftrightarrow} \Sigma^{y y}$ & $\Sigma^{x z} \stackrel{(-)}{\leftrightarrow} \Sigma^{y z}$\nn
\hline
$\Sigma^{x x}$ &
$\Sigma^{\circ y} \stackrel{(-)}{\leftrightarrow} \Sigma^{x z}$ & $\Sigma^{\circ z} \stackrel{(+)}{\leftrightarrow} \Sigma^{x y}$ &
$\Sigma^{y \circ} \stackrel{(-)}{\leftrightarrow} \Sigma^{z x}$ & $\Sigma^{z \circ} \stackrel{(+)}{\leftrightarrow} \Sigma^{y x}$\nn
$\Sigma^{x y}$ &
$\Sigma^{\circ x} \stackrel{(+)}{\leftrightarrow} \Sigma^{x z}$ & $\Sigma^{\circ z} \stackrel{(-)}{\leftrightarrow} \Sigma^{x x}$ &
$\Sigma^{y \circ} \stackrel{(-)}{\leftrightarrow} \Sigma^{z y}$ & $\Sigma^{z \circ} \stackrel{(+)}{\leftrightarrow} \Sigma^{y y}$\nn
$\Sigma^{x z}$ &
$\Sigma^{\circ x} \stackrel{(-)}{\leftrightarrow} \Sigma^{x y}$ & $\Sigma^{\circ y} \stackrel{(+)}{\leftrightarrow} \Sigma^{x x}$ &
$\Sigma^{y \circ} \stackrel{(-)}{\leftrightarrow} \Sigma^{z z}$ & $\Sigma^{z \circ} \stackrel{(+)}{\leftrightarrow} \Sigma^{y z}$\nn
$\Sigma^{y x}$ &
$\Sigma^{\circ y} \stackrel{(-)}{\leftrightarrow} \Sigma^{y z}$ & $\Sigma^{\circ z} \stackrel{(+)}{\leftrightarrow} \Sigma^{y y}$ &
$\Sigma^{x \circ} \stackrel{(+)}{\leftrightarrow} \Sigma^{z x}$ & $\Sigma^{z \circ} \stackrel{(-)}{\leftrightarrow} \Sigma^{x x}$\nn
$\Sigma^{y y}$ &
$\Sigma^{\circ x} \stackrel{(+)}{\leftrightarrow} \Sigma^{y z}$ & $\Sigma^{\circ z} \stackrel{(-)}{\leftrightarrow} \Sigma^{y x}$ &
$\Sigma^{x \circ} \stackrel{(+)}{\leftrightarrow} \Sigma^{z y}$ & $\Sigma^{z \circ} \stackrel{(-)}{\leftrightarrow} \Sigma^{x y}$\nn
$\Sigma^{y z}$ &
$\Sigma^{\circ x} \stackrel{(-)}{\leftrightarrow} \Sigma^{y y}$ & $\Sigma^{\circ y} \stackrel{(+)}{\leftrightarrow} \Sigma^{y x}$ &
$\Sigma^{x \circ} \stackrel{(+)}{\leftrightarrow} \Sigma^{z z}$ & $\Sigma^{z \circ} \stackrel{(-)}{\leftrightarrow} \Sigma^{x z}$\nn
$\Sigma^{z x}$ &
$\Sigma^{\circ y} \stackrel{(-)}{\leftrightarrow} \Sigma^{z z}$ & $\Sigma^{\circ z} \stackrel{(+)}{\leftrightarrow} \Sigma^{z y}$ &
$\Sigma^{x \circ} \stackrel{(-)}{\leftrightarrow} \Sigma^{y x}$ & $\Sigma^{y \circ} \stackrel{(+)}{\leftrightarrow} \Sigma^{x x}$\nn
$\Sigma^{z y}$ &
$\Sigma^{\circ x} \stackrel{(+)}{\leftrightarrow} \Sigma^{z z}$ & $\Sigma^{\circ z} \stackrel{(-)}{\leftrightarrow} \Sigma^{z x}$ &
$\Sigma^{x \circ} \stackrel{(-)}{\leftrightarrow} \Sigma^{y y}$ & $\Sigma^{y \circ} \stackrel{(+)}{\leftrightarrow} \Sigma^{x y}$\nn
$\Sigma^{z z}$ &
$\Sigma^{\circ x} \stackrel{(-)}{\leftrightarrow} \Sigma^{z y}$ & $\Sigma^{\circ y} \stackrel{(+)}{\leftrightarrow} \Sigma^{z x}$ &
$\Sigma^{x \circ} \stackrel{(-)}{\leftrightarrow} \Sigma^{y z}$ & $\Sigma^{y \circ} \stackrel{(+)}{\leftrightarrow} \Sigma^{x z}$
\end{tabular}
\caption{\label{Trotations}(Color Online)
Effect of $SU(4)$ rotations from Eqn.~(\ref{Egenerators}) 
by an arbitrary angle $\eta\in\mathbb{R}$ around the single axis specified 
by the first column.
For each transformation, only the displayed four pairs of axes are rotated 
into each other while the remaining axes are kept invariant.
The direction of the rotation is indicated by the sign and the order in 
every pair: 
For every pair in row $\Sigma^{\alpha\beta}$, an entry 
$\Sigma^{\gamma\delta} \stackrel{(\pm)}{\leftrightarrow} \Sigma^{\eta\lambda}$
denotes the identity
$e^{+\ii\eta\Sigma^{\alpha\beta}} \Sigma^{\gamma\delta} e^{-\ii\eta\Sigma^{\alpha\beta}} = 
\cos(\eta) \Sigma^{\gamma\delta} \pm \sin(\eta) \Sigma^{\eta\lambda}$.
The first six rotations do not mix between the uncoupled 
($\Sigma^{\circ x}$ to $\Sigma^{z \circ}$) and coupled 
($\Sigma^{x x}$ to $\Sigma^{z z}$) 
sector, whereas the remaining 9 rotations mix four axes from the coupled 
with four from the uncoupled block.
In particular, for $\eta=\pi/2$ we obtain a unitary transformation 
mapping coupled terms in a Hamiltonian to uncoupled ones and {\em vice versa}.
}
\end{table}
The observation that single-axis rotations around 
the axes from $\Sigma^{x x}$ to $\Sigma^{z z}$
may resolve couplings when the rotation angle $\eta=\pi/2$ yields a tool 
for the mapping of complicated (coupled) Hamiltonians 
towards simple (decoupled) ones.

For fixed-angle rotations $\alpha=\pi/2$ 
-- and in the following we will constrain ourselves to this case -- 
these mappings may also be represented graphically:
Identifying the $\sigma^x$, $\sigma^y$, and $\sigma^z$ operators with
squares, hexagons and circles, respectively, we may express local fields
by disconnected symbols, whereas many-body operators may be expressed by 
connected symbols.
The symbols may be grouped at different positions to indicate which 
qubit they are 
acting on and two qubits on which unitary operations act may be highlighted.
Note that we have not specified prefactors, which are simply transferred 
to the result.
Therefore, our results equally apply to models with different prefactors 
(e.g., in the presence of disorder).
For later reference we summarize the non-trivial action of the CZ gate
\bea\label{Ecz}
S_{ij}^{\rm CZ} \Sigma_{ij}^{\circ x} S_{ij}^{\rm CZ} 
&=& \Sigma_{ij}^{zx}\,,\qquad
S_{ij}^{\rm CZ} \Sigma_{ij}^{\circ y} S_{ij}^{\rm CZ} 
= \Sigma_{ij}^{zy}\,,\nn
S_{ij}^{\rm CZ} \Sigma_{ij}^{x\circ} S_{ij}^{\rm CZ} 
&=& \Sigma_{ij}^{xz}\,,\qquad
S_{ij}^{\rm CZ} \Sigma_{ij}^{y\circ} S_{ij}^{\rm CZ} 
= \Sigma_{ij}^{yz}\,,\nn
S_{ij}^{\rm CZ} \Sigma_{ij}^{xx} S_{ij}^{\rm CZ} 
&=& \Sigma_{ij}^{yy}\,,\qquad
S_{ij}^{\rm CZ} \Sigma_{ij}^{xy} S_{ij}^{\rm CZ} 
= -\Sigma_{ij}^{yx}\,,
\eea
where we remind the reader that the gate is its own inverse.
Obviously, the gate also commutes with itself when applied to different qubits and is symmetric, i.e., 
$S_{ij}^{\rm CZ} = S_{ji}^{\rm CZ}$.
Furthermore, we provide the action of the CNOT gate
\bea\label{Ecx}
S_{ij}^{\rm CX} \Sigma_{ij}^{\circ y} S_{ij}^{\rm CX} 
&=& \Sigma_{ij}^{zy}\,,\qquad
S_{ij}^{\rm CX} \Sigma_{ij}^{\circ z} S_{ij}^{\rm CX} 
= \Sigma_{ij}^{zz}\,,\nn
S_{ij}^{\rm CX} \Sigma_{ij}^{x\circ} S_{ij}^{\rm CX} 
&=& \Sigma_{ij}^{xx}\,,\qquad
S_{ij}^{\rm CX} \Sigma_{ij}^{y\circ} S_{ij}^{\rm CX} 
= \Sigma_{ij}^{yx}\,,\nn
S_{ij}^{\rm CX} \Sigma_{ij}^{xy} S_{ij}^{\rm CX} 
&=& \Sigma_{ij}^{yz}\,,\qquad
S_{ij}^{\rm CX} \Sigma_{ij}^{xz} S_{ij}^{\rm CX} 
= -\Sigma_{ij}^{yy}\,,
\eea
which is also its own inverse.
The other products of Pauli matrices -- not mentioned in Eqns.~(\ref{Ecz}) and~(\ref{Ecx}) --
are left invariant.
One could now in principle start from a known model and generate further 
models by arbitrarily applying two-qubit gates. 
To demonstrate the usefulness of our dualities, we will however rather 
try to map models with unknown properties to models with known ones.

\section{Discrete dualities for One-Dimensional Systems}

\subsection{Dualities of the Ising model in a transverse field}

The quantum Ising model in a transverse field
\bea\label{Eising}
H=-g \sum_{i=1}^N \sigma^x_{i}-J\sum_{i=1}^{N-1} \sigma^z_{i}\sigma^z_{i+1}
\eea
may be mapped for all $g$ and $J$ by a non-local
Jordan-Wigner transformation to a system of non-interacting 
fermions~\cite{pfeuty1970a}.
In the thermodynamic limit, it undergoes a quantum phase transition of 
second order at the critical point $J=g$ and
is the paradigmatic exactly solvable model within its universality class~\cite{kogut1979a}.
Its closed version (obtained by simply adding a $-J \sigma^z_N \sigma^z_1$ term) 
even admits a simple analytic diagonalization at finite sizes 
$N$~\cite{sachdev2000}.

\begin{figure}[ht]
\includegraphics[width=0.48\textwidth,clip=true]{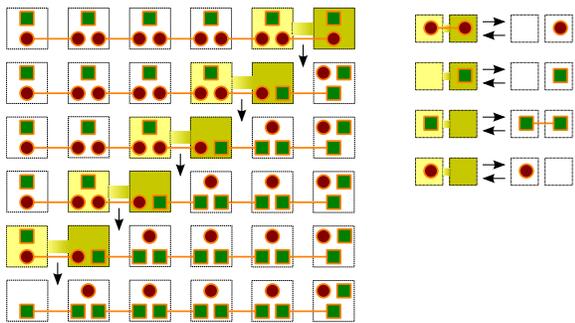}
\caption{\label{Fising_self}(Color Online)
Unitary mapping of the open Ising model with $N=6$ spins by a sequence 
of CNOT gates $S_{12}^{\rm CX}S_{23}^{\rm CX}S_{34}^{\rm CX}S_{45}^{\rm CX}S_{56}^{\rm CX}$ 
as in Eq.~(\ref{Ecx}).
Boxes in the background denote qubits, green squares and red circles 
denote $\sigma^x$ and $\sigma^z$ operators, respectively, and
orange connections denote many-body interactions.
Filled boxes denote the intended action of a CNOT gate~(\ref{Egatecx}) 
on the respective qubits,
which has been performed in each row below (compare legend on the right).
For example, the second row is obtained from the first by applying a $S_{56}^{\rm CX}$ gate.
Finally, applying local rotations for all spins demonstrates self-duality 
of the Ising model in a transverse field 
(up to boundary terms negligible in the large $N$-limit).
}
\end{figure}

First, by applying $SU(4)$ unitaries to the Ising model~(\ref{Eising}),
we find that it is self-dual in the infinite size limit $N\to\infty$,
via the unitary transformation $S^{\rm CX}_{12}\ldots S^{\rm CX}_{N-1,N}$.
This duality can be deduced algebraically from Eqns.~(\ref{Ecx}) or 
graphically from the diagrammatic representation in Fig.~\ref{Fising_self}.
Note that since the CNOT gates do not commute with each other, 
their order is relevant.
This duality fixes the phase transition point (if existent) 
to $J=g$, see also~\cite{doherty2009a} for a similar argument.
For finite chain lenghts however, one will map to an Ising model with 
modifications in the boundaries, see Fig.~\ref{Fising_self}.
When the prefactors are taken into account, we see that local field 
terms are mapped to ferromagnetic interactions
and {\em vice versa}, such that we obtain that Eq.~(\ref{Eising}) is dual to
\bea
H'=-J \sum_{i=2}^N \sigma^z_i - 
g \sum_{i=1}^{N-1} \sigma^x_i \sigma^x_{i+1} - g \sigma^x_N\,.
\eea
A similiar result can also be obtained from the Kramers-Wannier self 
duality~\cite{fradkin1978a,kogut1979a,peschel2004a,wolf2006a}.
{We also remark that such self-dualities can also be found in generalizations of the Ising model~\cite{wegner1971a,fradkin1979a}.}

Second, from properties of the CZ-gate~(\ref{Ecz}) it directly follows that 
the Ising model~(\ref{Eising}) is dual to the one-dimensional 
transverse-field cluster model Hamiltonian 
\bea
H'&=&-g \sum_{i=2}^{N-1}\sigma^z_{i-1}\sigma^x_{i}\sigma^z_{i+1}-
J\sum_{i=1}^{N-1}\sigma^z_{i}\sigma^z_{i+1}
-g \sigma^x_1 \sigma^z_2 -g \sigma^z_{N-1} \sigma^x_N
\eea
via the sequence $S^{\rm CZ}_{12}\ldots S^{\rm CZ}_{N-1,N}$, compare also 
\cite{schaller2008b}.
Here, as these gates commute, their order is not relevant and we do not 
provide a figure for brevity.

\begin{figure}[ht]
\includegraphics[width=0.48\textwidth,clip=true]{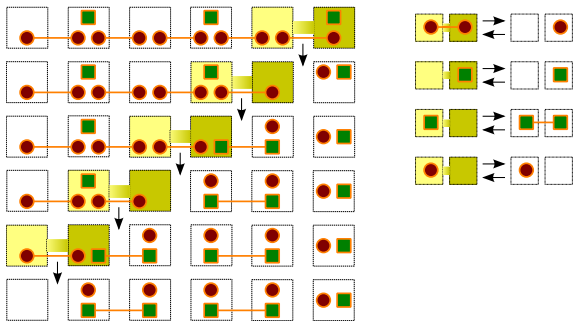}
\caption{\label{Fising_staggered}(Color Online)
A quantum Ising model with a staggered transverse field (top row) 
is sequentially
mapped to decoupled finite-dimensional subsystems (bottom row).
The empty box in the bottom row demonstrates that even with disorder, 
all eigenvalues of the corresponding Hamiltonian are two-fold degenerate.
Symbols and color coding have been chosen as in Fig.~\ref{Fising_self}.
}
\end{figure}

Third, a special version of the Ising model~(\ref{Eising}), 
where the local fields are only present at 
even sites (for simplicity we assume that $N$ is even)
\bea
H = -J \sum_{i=1}^{N-1} \sigma^z_i \sigma^z_{i+1} - 
g \sum_{i=1}^{N/2} \sigma^x_{2i}
\eea
can be easily mapped to decoupling two-qubit Ising
models with local fields, see Fig.~\ref{Fising_staggered}, 
by applying a sequence of CNOT operations in the same order.
Even more, one finds that after the transformation 
\bea
H'&=&\sum_{i=1}^{N/2-1} \left[-g \sigma^x_{2i} \sigma^x_{2i+1} - 
J \left(\sigma^z_{2i}+\sigma^z_{2i+1}\right)\right]
-J \sigma^z_N -g \sigma^x_N
\eea
there are no operators left acting on the first spin.
This automatically implies that even with disorder 
(different prefactors $g\to g_i$ and $J\to J_i$
for all operators~\cite{bunder1999a}) all eigenvalues are two-fold degenerate.

\subsection{The XZ Model}

As another example, the XY model without transverse field can be easily rotated into an XZ 
model by local transformations
\bea\label{Exzmodel}
H=\sum_{i=1}^{N-1}\left[ \alpha_i \sigma^z_i \otimes \sigma^z_{i+1} 
+ \beta_i \sigma^x_i \otimes \sigma^x_{i+1} \right]\,,
\eea
where we assume again for simplicity that $N$ is even.
This model can again be mapped to a quadratic fermionic Hamiltonian 
by means of a Jordan-Wigner transformation.
Afterwards, depending on the coefficients $\alpha_i$ and $\beta_i$ 
one could proceed with standard methods~\cite{bunder1999a} to map to free fermions.

\begin{figure}[ht]
\includegraphics[width=0.48\textwidth,clip=true]{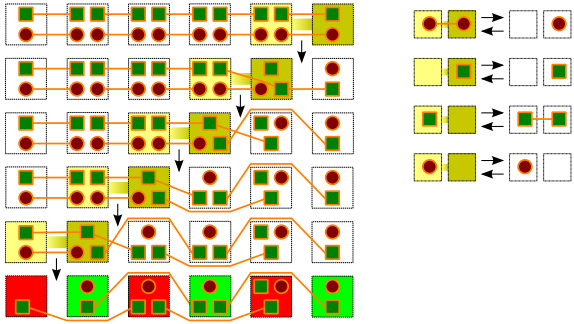}
\caption{\label{Fxz_ising}(Color Online)
Unitary Mapping between the XZ Model and two 
decoupled Ising models with (up to boundary effects) 
transverse fields by successive application of CNOT gates.
On the bottom row, there are no connections between even 
(light green) and odd (dark red) lattice sites.
Symbols and color coding have been chosen as in Fig.~\ref{Fising_self}.
}
\end{figure}

Here, we demonstrate that by a sequence of CNOT gates the XZ 
model~(\ref{Exzmodel}) is unitarily equivalent to two 
decoupled Ising models with transverse fields~\cite{mansson2013a}, 
separately defined on even and odd sites, respectively, 
which for brevity we only present graphically in Fig.~\ref{Fxz_ising}.

\subsection{One-Dimensional Cluster State Dualities}

From the properties of the CZ gate~(\ref{Ecz}), it is obvious that the
Hamiltonian encoding at $g=0$~\cite{briegel2001a,bacon2010a} the one-dimensional cluster state in its ground state 
\bea
H&=&-J \sigma^x_1 \sigma^z_2 
-J \sum_{i=2}^{N-1} \sigma^z_{i-1}\sigma^x_{i}\sigma^z_{i+1} 
-J\sigma^z_{N-1}\sigma^x_N
- g \sum_{i=1}^N \sigma^z_{i}
\eea
is dual to a Hamiltonian for non-interacting qubits
\bea
H'=-J \sum_{i=1}^N \sigma^x_{i}-g \sum_{i=1}^N \sigma^z_{i}
\eea
via the unitary transformation $S^{\rm CZ}_{12}\ldots S^{\rm CZ}_{n-1,n}$ 
(not shown, but see also Fig.~\ref{Fclusterstate}).
As the fundamental energy gap $2\sqrt{g^2+J^2}$ above the ground state does 
not scale with the
system size $N$, this enables the adiabatic preparation of one-dimensional 
cluster states from the 
$z$-polarized phase by linearly interpolating from $J=0$ to $g=0$ in constant time, 
independent of the system size~\cite{kalis2012a}.
Compared to the conventional cluster state preparation~\cite{briegel2001a}, 
this has the additional advantage that the desired evolution
can be encoded in the unique and robust ground state.

On the other hand, when the local field points towards another direction, 
the Hamiltonian
\bea
H&=&-J \sigma^x_1 \sigma^z_2 
-J \sum_{i=2}^{N-1} \sigma^z_{i-1}\sigma^x_{i}\sigma^z_{i+1}
-J \sigma^z_{N-1} \sigma^z_N 
-g\sum_{i=1}^N \sigma^x_{i}
\eea
is self-dual via $S^{\rm CZ}_{12}\ldots S^{\rm CZ}_{n-1,n}$~\cite{doherty2009a} 
(not shown).
In fact, this model can be mapped to two decoupled
Ising models in one dimension~\cite{doherty2009a}, and
exhibits a second order quantum phase transition, which is associated with an 
inverse scaling of the minimum energy gap with the system size $N$ as 
$g_{\rm min} = \ord\{1/N\}$~\cite{pachos2004a}.
Similar mappings to transverse-field Ising chains exist for generalizations~\cite{lahtinen2015a}.

\section{Two-Dimensional Systems}

\subsection{Two-Dimensional Cluster State Dualities}

Universal measurement-based quantum 
computation cannot be achieved with the one-dimensional cluster state.
{The two-dimensional cluster state however provides sufficient resources for this task~\cite{raussendorf2001a,raussendorf2003a}.
It can also be used for the preparation of topological order~\cite{brown2011a}.}
In order to estimate the preparation complexity when the two-dimensional 
cluster state
is adiabatically prepared by slowly deforming a control parameter, 
we consider the model
\bea
\label{2D-cluster-1}
H = 
- J \sum_{\mu} \left[\bigotimes_{\nu \sim \mu}\sigma^z_\nu\right]\sigma^x_\mu 
- g \sum_\mu \sigma^x_\mu\,,
\eea
where $\mu$ involves all sites of a lattice and $\nu \sim \mu$ denotes all 
neighbors of $\mu$.
The CZ-gate -- note its symmetry in Eqns.~(\ref{Ecz}) -- 
when applied to all links in the lattice immediately demonstrates
self-duality of the model in any dimension.
In particular for two dimensions, the model can be 
mapped~\cite{doherty2009a} to the Xu-Moore compass model~\cite{xu2004a,xu2005a,nussinov2005a}
exhibiting a second order quantum phase transition when $J=g$.
The shrinking of the energy gap associated with the quantum phase 
transition does not only lead to a scaling of the adiabatic preparation
time with the system size, but also to an increased vulnerability with 
respect to thermal excitations~\cite{childs2001a,mostame2010a}.

\begin{figure}[ht]
\includegraphics[width=0.45\textwidth,clip=true]{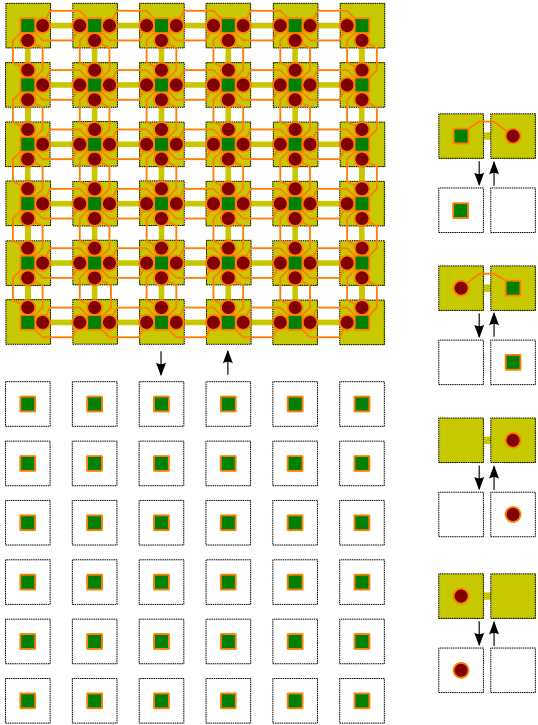}
\caption{\label{Fclusterstate}(Color Online)
The CZ gate -- applied to all links on the lattice -- can be used to 
decouple the 5-body interactions in the 
Hamiltonian for the cluster state completely.
Since it commutes with the $\sigma^z_i$ operators, it will not affect 
any additional local fields in $z$-direction (not shown).
Symbol and Color Coding as in Fig.~\ref{Fising_self}.
}
\end{figure}

In contrast, the Hamiltonian 
\bea
\label{2D-cluster-2}
H = 
-J \sum_{\mu} \left[\bigotimes_{\nu \sim \mu}\sigma^z_\nu\right]\sigma^x_\mu 
-g \sum_\mu \sigma^z_\mu\,,
\eea
where only the local field is pointing into a different direction, 
is dual to the completely decoupled one
\bea
H' = - J \sum_{\mu} \sigma^x_\mu - g \sum_\mu \sigma^z_\mu\,,
\eea
for clarity we illustrate this for $g=0$ only in Fig.~\ref{Fclusterstate}.

Consequently, the energy gap above the ground state of the 
Hamiltonian~(\ref{2D-cluster-2}) is independent of the system size -- 
in contrast to the Hamiltonian~(\ref{2D-cluster-1}). 
As a result, this facilitates the robust adiabatic preparation of the 
two-dimensional cluster states by linear interpolation.
This straightforwardly generalizes to cluster states on lattices in arbitrary dimensions.
We note here that adiabatic rotation~\cite{siu2007a} also enables for a preparation of the two-dimensional 
cluster state in constant time, which however requires very complex interpolation paths.

\subsection{From 1D to 2D: Hexagonal Lattice}

\begin{figure}[ht]
\includegraphics[width=0.45\textwidth,clip=true]{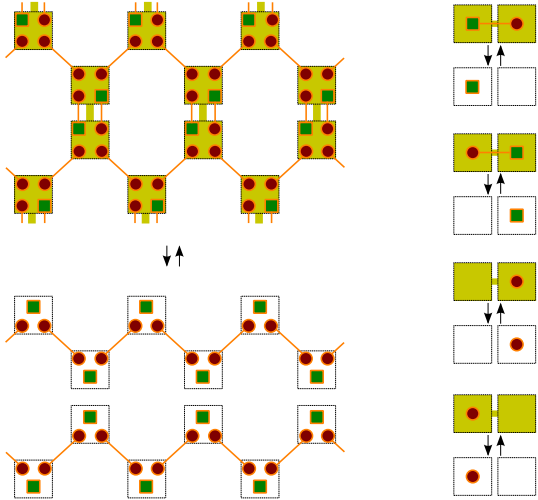}
\caption{\label{Fising2d}(Color Online)
Parallel Ising chains (bottom left) may be mapped to a hexagonal lattice 
(top left) by using CZ mappings on every site
either with the upper or the lower chain in an alternating fashion.
}
\end{figure}

In order to consider a different lattice geometry, let us
consider a number of parallel but mutually uncoupled Ising chains as 
depicted in the bottom left panel of Fig.~\ref{Fising2d}.
Now, by applying a CZ gate to each spin of the Ising chain with the 
spins of the neighbouring chains in an
alternating fashion, it is straightforward to see that this will induce 
a coupling.
The resulting two-dimensional model shown in the top left panel will 
inherit its critical behaviour from the 1d Ising chains.
It is interesting to note that there will be long-range entanglement 
in the longitudinal direction but not 
in the transverse direction, since all CZ gates commute.

%

\subsection{From 1D to 2D: Plaquette Hamiltonians}

{In two-dimensional spin systems subject to plaquette-shaped many-body interactions, 
exciting new phenomena may arise.
Plaquette operators are basic building blocks of the toric code~\cite{kitaev2003a}, which -- as well as
with related models~\cite{xu2004a,xu2005a,trebst2007a,hamma2008a,vidal2009b,schmidt2013a} -- supports topologically 
ordered phases.}
\begin{figure}[ht]
\includegraphics[width=0.45\textwidth,clip=true]{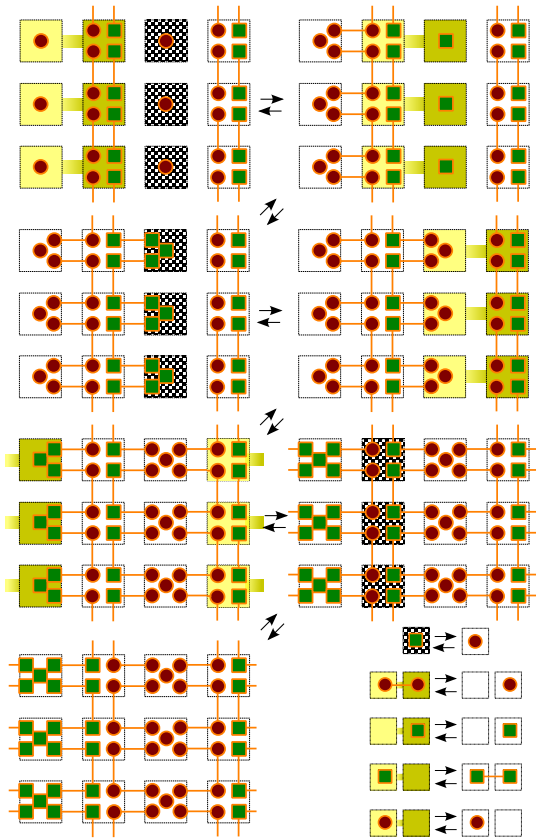}
\caption{\label{Fplaquette1}(Color Online)
Vertical XZ chains together with local fields (top left $4\times 3$ lattice) 
may be mapped to a Hamiltonian with monochromatic $X$ and $Z$ plaquettes and 
non-homogeneous local fields (bottom left $4\times 3$ lattice).
This requires a sequence of local rotations (checkerboard sites) and CNOT gates.
}
\end{figure}

It is of course also possible to use the CNOT gate, for example, to generate 
two-dimensional models from known one-dimensional ones.
However, this gate will not commute with itself, such that the order 
at which such gates are applied is relevant.
Consider for example a collection of one-dimensional parallel XZ models 
as depicted in the top left panel of
Fig.~\ref{Fplaquette1}.
Between the XZ chains, we have placed local fields, 
such that the total lattice has a simple cubic structure.
Using the sites with local field as a control qubit, 
a column of applied CNOT gates generates
monochromatic plaquette operators in $z$-direction.
To use the CNOT gate again, we perform a local rotation 
(checkerboard background).
Then, repeating the application of CNOT gates -- 
moved one lattice spacing to the right -- 
will generate monochromatic plaquette operators in $x$-direction.
This sequence of local rotations and CNOT gates can be continued 
until the boundary is reached or
-- in case of closed boundary conditions -- the already existing 
plaquettes are met.
The final local rotation in Fig.~\ref{Fplaquette1} is only 
performed to generate monochromatic plaquettes {that are usually used in compass models}.
As the CNOT gate does not commute with itself, the coupling may 
now create long-range entanglement also in the transverse direction.

\subsection{Braiding Plaquettes into netting wires}

\begin{figure}[ht]
\includegraphics[width=0.45\textwidth,clip=true]{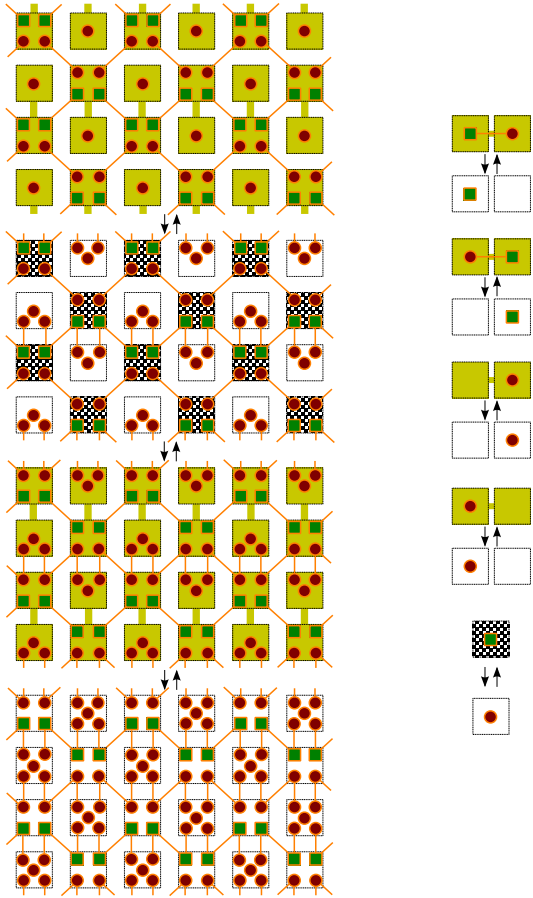}
\caption{\label{Fplaquette2}(Color Online)
A ferromagnetic netting wire with local fields (top panel) 
may be unitarily transformed
into plaquettes with local fields (bottom panel) by a sequence of 
CZ gates and local rotations.
}
\end{figure}

Finally, we note that plaquette operators in a two-dimensional 
lattice can also be created with CZ gates.
As an example consider the netting wire constructed from 
ferromagnetic interactions
in $x$ and $z$ directions in the top panel of Fig.~\ref{Fplaquette2}.
In the mid of the diamonds we place local fields, which we can locally 
rotate into $z$ direction
without loss of generality.
Obviously, there will be no long-distance entanglement in the 
eigenstates of the Hamiltonian
between two spins on a local field site.
It is straightforward to see that a sequence of CZ gates in 
combination with local rotations
in appropriate order will generate plaquette operators throughout the 
lattice.

%


\section{Summary and Outlook}

Using the $SU(4)$ algebra, we have demonstrated that one can map many 
seemingly complicated models to known ones and {\em vice versa}.
The unitary mappings enabled us to demonstrate dualities for spin systems, 
which may be used to identify the position of critical points (self-dualities) 
or to draw conclusions on the spectrum.
In some cases, the transformations even allowed to map to decoupled 
finite-dimensional subsystems and thereby provide the complete diagonalization 
of the original model.
As an interesting application, our method shows that the two-dimensional 
cluster state (useful for universal measurement-based
quantum computation) can be efficiently prepared adiabatically using only
linear interpolation, i.e., the Hamiltonian~(\ref{2D-cluster-2}).

Our list of examples is of course by far not complete and it is certainly 
interesting to find more dualities.
Note also that we have constrained ourselves to discrete gates, 
where the rotation angle is fixed, to facilitate
a graphical representation.
This however is not a fundamental limitation.
We hope that the transformation table of $SU(4)$ may aid in tailoring 
appropriate unitary transformations for other spin systems, too.


\section{Acknowledgements}

G. S. gratefully acknowledges stimulating discussions with 
V. M. Bastidas, T. Brandes and J. Eisert and 
financial support by the DFG (SCHA 1646/3-1, BRA 1528/7-1).



\begin{thebibliography}{9999}

\bibitem{nielsen2000}
M. A. Nielsen and I. L. Chuang,
{\em Quantum Computation and Quantum Information},
Cambridge University Press, Cambridge (2000).

\bibitem{pfeuty1970a}
P. Pfeuty, 
Ann. Phys. {\bf 57}, 79 (1970).

{\bibitem{kogut1979a}
J. B. Kogut, 
Reviews of Modern Physics {\bf 51}, 659 (1979).
}

\bibitem{sachdev2000}
S. Sachdev,
{\em Quantum Phase Transitions},
Cambridge University Press, Cambridge (2000).

\bibitem{doherty2009a}
A. C. Doherty and S. D. Bartlett,
Physical Review Letters {\bf 103}, 020506 (2009).

{\bibitem{fradkin1978a}
E. Fradkin and L. Leonard, 
Physical Review D {\bf 17}, 2637 (1978).
}

\bibitem{peschel2004a}
I. Peschel,
Journal of Statistical Mechanics: Theory and Experiment {\bf 04}, P12005 (2004).

\bibitem{wolf2006a}
M. M. Wolf, G. Ortiz, F. Verstraete, and J. I. Cirac,
Physical Review Letters {\bf 97}, 110403 (2006).

{\bibitem{wegner1971a}
F. J. Wegner, 
Journal of Mathematical Phyics {\bf 12}, 2259 (1971).
}

{\bibitem{fradkin1979a}
E. Fradkin and S. H. Shenker,
Physical Review D {\bf 19}, 3682 (1979).
}

\bibitem{schaller2008b}
G. Schaller, 
Phys. Rev. A {\bf 78}, 032328 (2008).

\bibitem{bunder1999a}
J. E. Bunder and R. H. McKenzie,
Physical Review B {\bf 60}, 344 (1999).

\bibitem{mansson2013a}
T. M\aa{}nsson, V. Lahtinen, J. Suorsa, and E. Ardonne,
Physical Review B {\bf 88}, 041403 (2013).

\bibitem{briegel2001a}
H. J. Briegel and R. Raussendorf,
Physical Review Letters {\bf 86}, 910 (2001).

\bibitem{bacon2010a}
D. Bacon and S. T. Flammia,
Physical Review A {\bf 82}, 030303 (2010).

{\bibitem{kalis2012a}
H. Kalis, D. Klagges, R. Or\'us, and K. P. Schmidt,
Physical Review A {\bf 86}, 022317 (2012).
}

\bibitem{pachos2004a}
J. K. Pachos and M. B. Plenio
Physical Review Letters {\bf 93}, 056402 (2004).

\bibitem{lahtinen2015a}
V. Lahtinen and E. Ardonne,
Physical Review Letters {\bf 115}, 237203 (2015).

\bibitem{raussendorf2001a}
R. Raussendorf and H. J. Briegel,
Physical Review Letters {\bf 86}, 5188 (2001).

\bibitem{raussendorf2003a}
R. Raussendorf, D. E. Browne, and H. J. Briegel,
Physical Review A {\bf 68}, 022312 (2003).

{\bibitem{brown2011a}
B. J. Brown, W. Son, C. V. Kraus, R. Fazio, and V. Vedral,
New Journal of Physics {\bf 13}, 065010 (2011).
}

\bibitem{xu2004a}
C. Xu and J. E. Moore, 
Physical Review Letters {\bf 93}, 047003 (2004).

{\bibitem{xu2005a}
C. Xu and J. E. Moore,
Nuclear Physics B {\bf 716}, 487 (2005).
}

{\bibitem{nussinov2005a}
Z. Nussinov and E. Fradkin,
Physical Review B {\bf 71}, 195120 (2005).
}

\bibitem{childs2001a}
A. M. Childs, E. Farhi, and J. Preskill,
Physical Review A {\bf 65}, 012322 (2001).

\bibitem{mostame2010a}
S. Mostame, G. Schaller, and Ralf Sch\"utzhold
Physical Review A {\bf 81}, 032305 (2010).

\bibitem{siu2007a}
M. S. Siu,
Physical Review A {\bf 75}, 062337 (2007).

{\bibitem{kitaev2003a}
A. Yu. Kitaev, 
Annals of Physics (N. Y.) {\bf 303}, 2 (2003).
}

{\bibitem{trebst2007a}
S. Trebst, P. Werner, M. Troyer, K. Shtengel, and C. Nayak,
Physical Review Letters {\bf 98}, 070602 (2007).
}

{\bibitem{vidal2009b}
J. Vidal, R. Thomale, K. P. Schmidt, and S. Dusuel,
Physical Review B {\bf 80}, 081104 (2009).
}

{\bibitem{schmidt2013a}
K. P. Schmidt,
Physical Review B {\bf 88}, 035118 (2013).
}

{\bibitem{hamma2008a}
A. Hamma and D. A. Lidar,
Physical Review Letters {\bf 100}, 030502 (2008).
}

\end{thebibliography}
\end{document}